\documentclass[iop,revtex4.1]{emulateapj}

\usepackage{natbib}
\usepackage{amsmath}
\usepackage{color}
\usepackage{url}
\usepackage{ulem}
\usepackage{multirow}

\bibpunct{(}{)}{;}{a}{}{,}

\def\apjl{ApJ}%
\def\apjs{ApJS}%
\def\ao{Appl.~Opt.}%
\def\aap{A\&A}%

\shorttitle{ {\it Suzaku} observation of IRAS\,00521$-$7054}
\shortauthors{Ricci et al.}

\begin{document}


\title{{\it Suzaku} observation of IRAS\,00521$-$7054, a peculiar type-II AGN with 

a very broad feature at 6\,keV}


%

\author{C. Ricci\altaffilmark{1,*}, F. Tazaki\altaffilmark{1,2}, Y. Ueda\altaffilmark{1}, S. Paltani\altaffilmark{3},  R. Boissay\altaffilmark{3}, Y. Terashima\altaffilmark{4}
}

\altaffiltext{1}{Department of Astronomy, Kyoto University, Oiwake-cho, Sakyo-ku, Kyoto 606-8502, Japan.}
\altaffiltext{2}{Mizusawa VLBI Observatory, National Astronomical Observatory of Japan, Mitaka, Tokyo 181-8588, Japan.}
\altaffiltext{3}{Department of Astronomy, University of Geneva, ch. d'Ecogia 16, 1290 Versoix, Switzerland.}
\altaffiltext{4}{Department of Physics, Ehime University, Matsuyama, 790-8577, Japan.}

\altaffiltext{*}{ricci@kusastro.kyoto-u.ac.jp}

\begin{abstract}
IRAS\,00521$-$7054 is the first Seyfert\,2 in which the presence of an extremely large Fe\,K$\alpha$ line has been claimed. We report here on the analysis of a 100\,ks {\it Suzaku} observation of the source. We confirm the existence of a very strong excess over the power-law X-ray continuum at $E\sim6$\,keV ($EW\simeq 800$\,eV), extending down to $\sim 4.5$\,keV, and found that the X-ray spectrum of the source can be explained by two different models.
i) An absorption scenario, in which the X-ray source is obscured by two fully-covering ionized absorbers, with a strong reflection component from neutral material ($R\sim 1.7$), a black body component and four narrow Gaussian lines (corresponding to Fe K$\alpha$, Fe K$\beta$, Fe\,\textsc{xxv} and Fe \textsc{xxvi}).
ii) A reflection scenario, in which the X-ray spectrum is dominated by an obscured ($\log N_{\rm\,H}\sim 22.9$) blurred reflection produced in an ionized disk around a rotating supermassive black hole with a spin of $a \geq 0.73$, and affected by light-bending ($R\sim 2.7$), plus two narrow Gaussian lines (corresponding to Fe K$\alpha$ and Fe K$\beta$).
The narrow Fe K$\alpha$ and K$\beta$ lines are consistent with being produced by ionized iron, and in particular by Fe\,\textsc{xiv}--Fe\,\textsc{xvi} and Fe\,\textsc{xii}--Fe\,\textsc{xvi} for the absorption and reflection scenario, respectively.
While the X-ray continuum varies significantly during the observation, the intensity of the broad feature appears to be constant, in agreement with both the absorption and reflection scenarios. For both scenarios we obtained a steep power-law emission ($\Gamma\sim 2.2-2.3$), and we speculate that the source might be an obscured narrow-line Seyfert\,1. 

\end{abstract}



\section{Introduction}
Iron emission lines are one of the most prominent features in the X-ray spectra of Active Galactic Nuclei (AGN), and are the most important tracers of the material surrounding the accreting supermassive black hole (SMBH). Most AGN show evidence of a narrow Fe\,K$\alpha$ line at 6.4\,keV, likely produced by reflection from neutral material (i.e. less ionized than Fe\,\textsc{xviii}, e.g., \citealp{Makishima:1986ve,Yaqoob:2001fv}) in the broad line region (BLR, e.g., \citealp{Bianchi:2008bs}) or in the molecular torus (e.g., \citealp{Shu:2010zr,Ricci:2014fk}).
The idea that the primary X-ray continuum reprocessed in the innermost part of the accretion disk might give rise to iron lines broadened by Doppler motion and relativistic effects was first proposed by \cite{Fabian:1989ve}. Few years later, using {\it ASCA}, \cite{Mushotzky:1995ve} and \cite{Tanaka:1995qf} found the first evidence of the existence of such broad Fe\,K$\alpha$ lines in the X-ray spectra of NGC\,5548, IC\,4329A and MCG$-$6$-$30$-$15. With the advent of {\it XMM-Newton} and {\it Suzaku}, broad lines have been detected in about $30-50\%$ of local bright unabsorbed AGN (e.g., \citealp{de-La-Calle-Perez:2010fk}, \citealp{Patrick:2012qf}). The ratio between the flux of the broad and the narrow component of the Fe K$\alpha$ line in the 6.35--6.45\,keV region is expected to be $\leq 25\%$, except in the case of truncated disks, as the disk emission is less blurred and more flux is emitted around the core of the line, or in the presence of light-bending \citep{Ricci:2014fv}. 
A possible alternative explanation to the existence of relativistic broad iron lines is that broad features are artificially created by the distortion of the X-ray continuum created by ionized absorbers (\citealp{Turner:2009fk} and references therein, \citealp{Miyakawa:2012vn}).

While they are rather common in type\,1-1.5 AGN, broad Fe\,K$\alpha$ lines have so far been found in only a few obscured objects:  the Seyfert\,2s IRAS\,18325$-$5926 \citep{Iwasawa:1996bh,Lobban:2014zr} and MCG$-$5$-$23$-$16  \citep{Dewangan:2003ys}, the Seyfert\,1.9 NGC\,2992 \citep{Shu:2010vn}, and the Seyfert\,1.8s IRAS\,13197$-$1627 \citep{Miniutti:2007fk} and NGC\,1365 \citep{Risaliti:2013ly}. Another absorbed AGN showing evidence of a broad Fe\,K$\alpha$ line is the obscured narrow-line Seyfert\,1 (NLS1) galaxy NGC\,5506 \citep{Guainazzi:2010kx}. 
Recently \cite{Tan:2012fk}, analysing two $\sim10$\,ks {\it XMM-Newton} observations of the Seyfert\,2 IRAS\,00521$-$7054 (z=0.0689), claimed to have detected a very strong and broad (with an equivalent width $EW=1.3\pm0.3\rm\,keV$) Fe\,K$\alpha$ line. \cite{Tan:2012fk} found that, to reproduce this large feature, a maximally rotating (with a spin of $a=0.97^{+0.03}_{-0.13}$) SMBH is needed.
However, due to the low signal to noise ratio of the {\it XMM-Newton} observations, they could not clearly distinguish between the scenario in which the line is relativistically broadened, and that in which the X-ray source is obscured by partially covering neutral absorbers.
We report here the study of a 100\,ks {\it Suzaku} follow-up observation of IRAS\,00521$-$7054, aimed at understanding the characteristics and the nature of its broad feature at 6\,keV.

\clearpage

\section{Data reduction}

\subsection{{\it Suzaku}}\label{sect:suzakudatared}
The {\it Suzaku} observatory \citep{Mitsuda:2007dq} carries on board a set of X-ray CCD cameras, the X-ray Imaging Spectrometer (XIS, \citealp{Koyama:2007cr}), and the non-imaging Hard X-ray Detector (HXD, \citealp{Takahashi:2007nx}), composed of Si positive intrinsic negative (PIN) photodiodes and Gadolinium Silicon Oxide (GSO) scintillation counters. XIS is composed of three operational cameras, the front-illuminated (FI) XIS\,0 and XIS\,3, and the back-illuminated (BI) XIS\,1 (hereafter BI-XIS).

IRAS\,00521$-$7054 was observed by {\it Suzaku} on 2013, May 31st (observation ID\,708005010) at the ``XIS nominal" pointing position. Each XIS camera had net exposure of 103\,ks, while the exposure of HXD/PIN was 87.3\,ks.
The light curves and spectra were extracted using HEAsoft version 6.12 and {\it Suzaku} CALDB version 20130916.

For each of the three XIS cameras we extracted light curves and spectra from XIS cleaned events, processed by the {\it Suzaku} team, using a circular region with a radius of 1.7\,arcmin centred on the source. The background was taken from a source-free annulus region centred at the source peak, with an internal and external radius of 3.5 and 5.7\,arcmin, respectively. We excluded from the background a circular region of 55\,arcsec around a possible very faint X-ray source.  We generated the ancillary response matrices (ARFs) and the detector response matrices (RMFs) using the \texttt{xisrmfgen} and \texttt{xissimarfgen} tasks \citep{Ishisaki:2007oq}, respectively. The spectra and light-curve obtained by XIS\,0 and XIS\,3, were merged (hereafter FI-XIS) using \texttt{mathpha}, \texttt{addrmf} and \texttt{addarf}.

The flux of IRAS\,00521$-$7054 is below the detection threshold of HXD/GSO, therefore we only analyzed HXD/PIN data. 
The HXD/PIN spectrum was extracted from the cleaned event files, and then corrected for dead time using \texttt{hxddtcor}. We adopted the response file corresponding to the 11th epoch of {\it Suzaku} observations and to the observation mode (\texttt{ae\_hxd\_pinxinome11\_20110601.rsp}). We produced the background spectra using the latest release of the tuned non-X-ray background (NXB) event files provided by the HXD team, which takes into account the 3--4\% underestimation of the NXB flux for observations carried out after July 31st 2012\footnote{http://www.astro.isas.jaxa.jp/suzaku/analysis/hxd/pinnxb/tuned/}. We added to the background spectrum that of the cosmic X-ray background (CXB) by using the formula of \cite{Gruber:1999tg}. We found that the PIN spectrum is consistent with the systematic error of the NXB (3\% of the NXB level, \citealp{Fukazawa:2009kl}), and therefore we did not use the data for the spectral analysis.
The 16--60\,keV PIN count-rate of the source before NXB subtraction is $0.211\pm 0.002\rm\,ct\,s^{-1}$, which, assuming a power-law emission, corresponds to a flux of $1.37\times 10^{-10}\rm\,erg\,cm^{-2}\,s^{-1}$. Consequently, taking into account the 3\% systematic uncertainty of the NXB, we can place an upper limit of $F_{15-60\rm\,keV}\lesssim 4.1\times 10^{-12}\rm\,erg\,cm^{-2}\,s^{-1}$ on the 15--60\,keV flux of IRAS\,00521$-$7054.

\begin{figure}[t!]
\centering
\centering
\includegraphics[width=9cm]{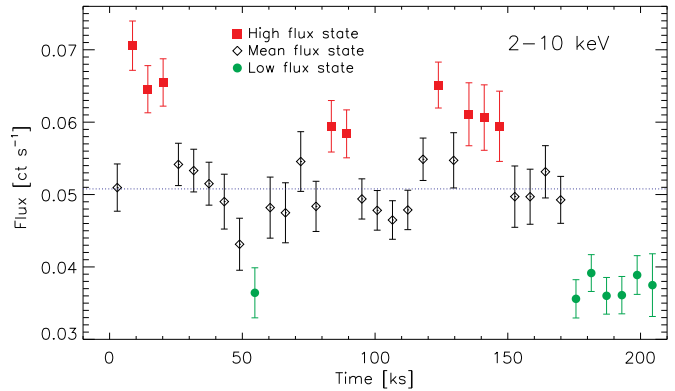}
  \caption{FI-XIS background subtracted 2--10\,keV light curve of IRAS\,00521$-$7054 in 5760\,s orbital bins. The horizontal dotted line represents the average flux of the observation. The red squares, black diamonds and green circles represent the high, mean and low flux state, respectively (see Sect.\,\ref{sect:lc} and Sect.\,\ref{sect:timeresolved}).}
\label{fig:xislc}
\end{figure}

\bigskip
\bigskip

\subsection{{\it XMM-Newton}}
We reanalysed the two {XMM-Newton} observations of IRAS\,00521$-$7054 discussed by \cite{Tan:2012fk}, taking into account the data obtained by the PN \citep{Struder:2001uq} and MOS \citep{Turner:2001fk} cameras on board {\it XMM-Newton}. The two observations were carried out on March\,22nd and April\,22nd 2006 (PI F. Nicastro).
The original data files (ODFs) were reduced using the {\it XMM-Newton} Standard Analysis Software (SAS) version 12.0.1 \citep{Gabriel:2004fk}, and the raw PN and MOS data files were then processed using the \texttt{epchain} and \texttt{emchain} tasks, respectively. 

For each observation we checked the background light curve in the 10--12 keV energy band for PN, and above 10\,keV for MOS, in order to filter the exposures for periods of high background activity. 
The threshold was set to $0.5\rm\,ct\,s^{-1}$ for the three detectors. Only patterns corresponding to single and double events (PATTERN~$\leq 4$) were selected for PN, and corresponding to single, double, triple and quadruple events for MOS (PATTERN~$\leq 12$). 
For the first observation (ID 0301150101) the net exposures were 7.8, 10.2 and 9.6\,ks for PN, MOS1 and MOS2, respectively. The source spectra were extracted from the final filtered event list using circular regions centred on the object, with a radius of 17.5, 14 and 14\,arcsec for PN, MOS1 and MOS2, respectively.
For the second observation (ID 0301151601) the exposures were 10.4 (PN), and 13.3\,ks (MOS1 and MOS2), while the radius was 22.5, 14 and 14\,arcsec for PN, MOS1 and MOS2, respectively. In both observations and for all cameras the background was estimated from circular regions with a radius of 40\,arcsec located on the same CCD of the source, where no other source was present. No pile-up was detected in the two observations for any of the three cameras. The ARFs and RMFs were created using the \texttt{arfgen} and \texttt{rmfgen} tasks, respectively.

\begin{figure*}[t!]
\centering
\begin{minipage}[!b]{.48\textwidth}
\centering
\includegraphics[width=9cm]{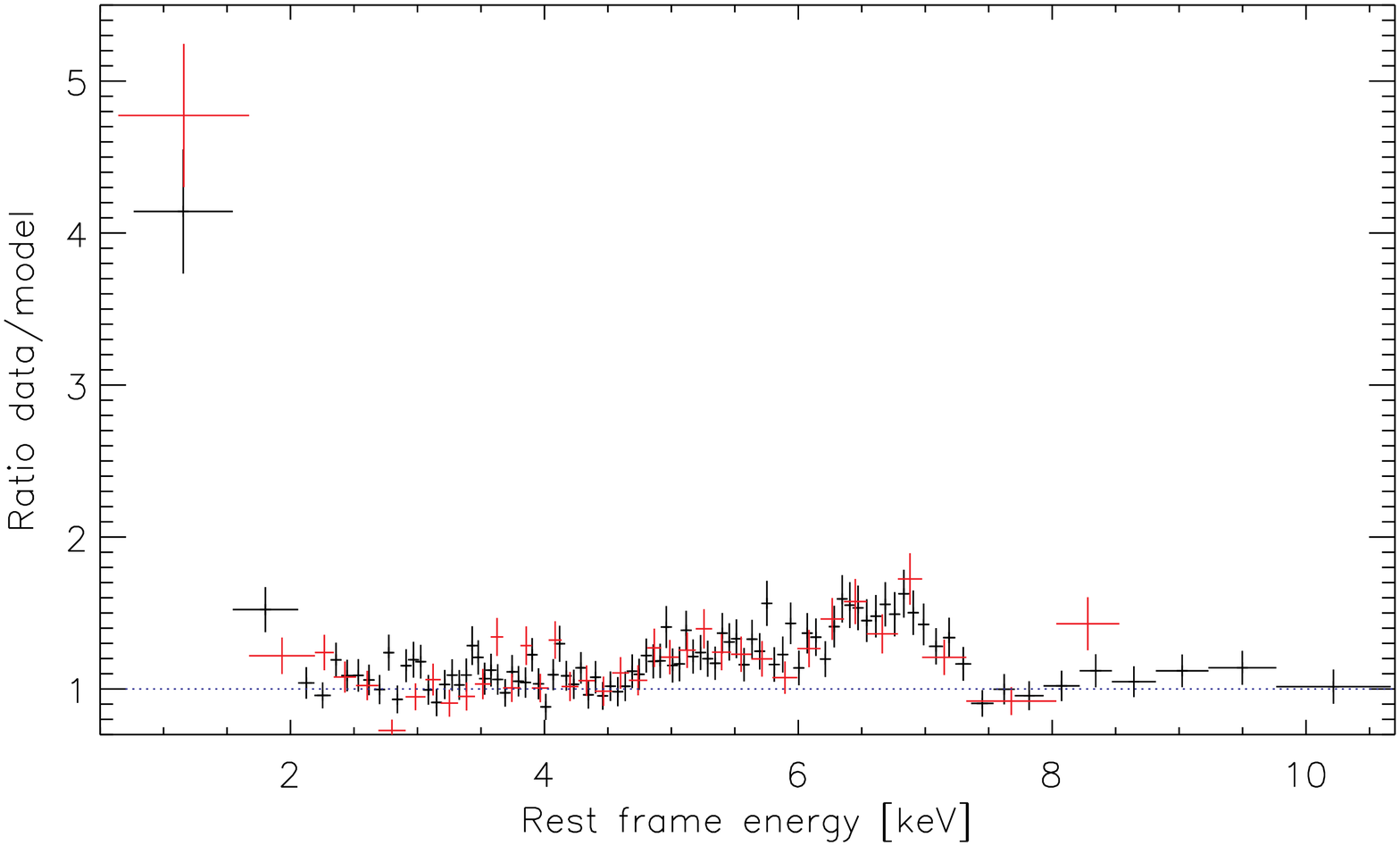}\end{minipage}
\hspace{0.05cm}
\begin{minipage}[!b]{.48\textwidth}
\centering
\includegraphics[width=9cm]{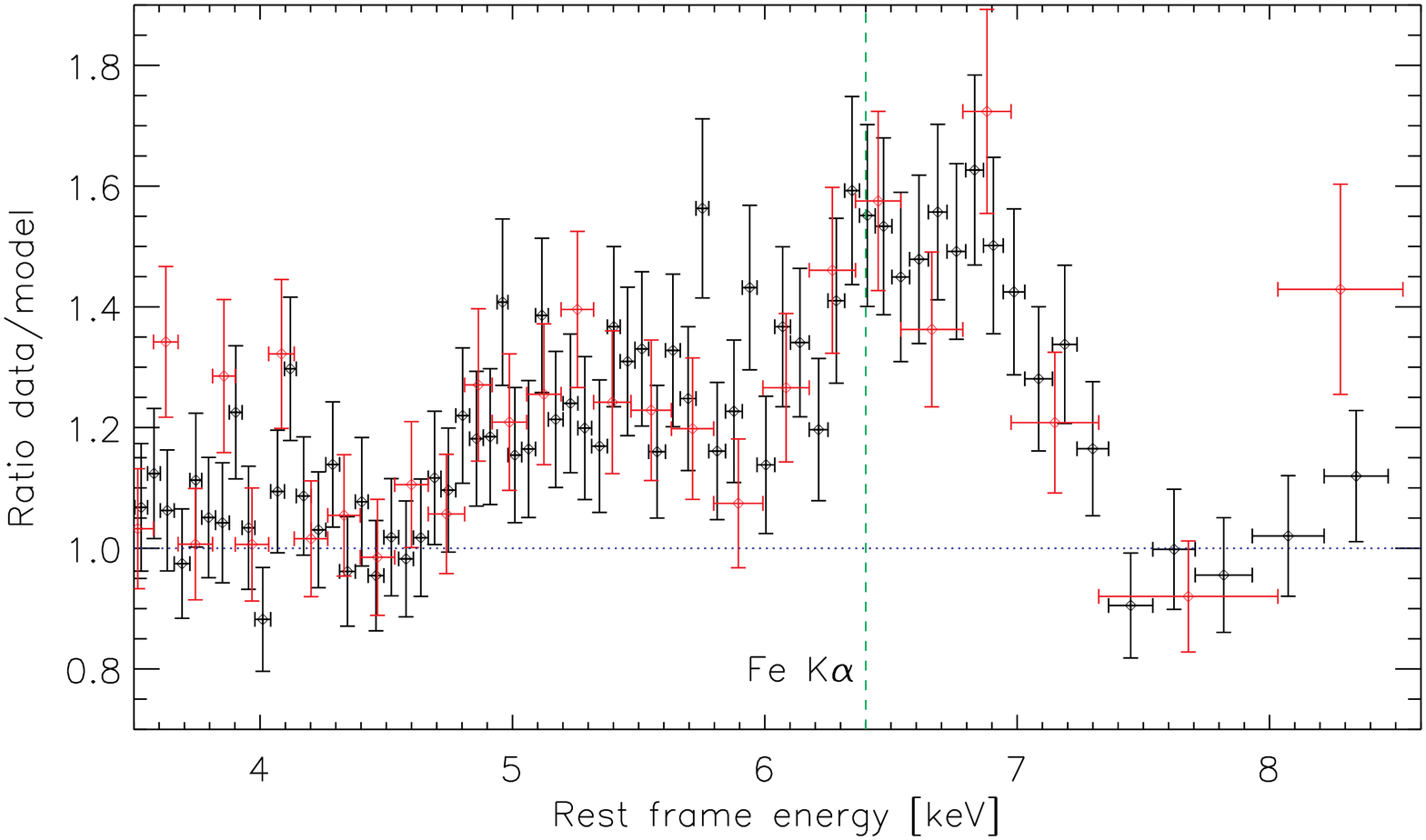}\end{minipage}
 \begin{minipage}[t]{1\textwidth}
  \caption{{\it Left panel:} ratio between {\it Suzaku}/XIS data and the fit to the spectrum obtained by applying a simple absorbed power-law model (\texttt{TBABS$_{\mathrm{Gal}}\times$ZTBABS$\times$ZPOW}) to the data in the 2--4.5\,keV and 7.5--10\,keV range (in the rest frame of the source). The black and red points represent the data obtained by FI-XIS and BI-XIS, respectively. {\it Right panel}: zoom in the 3.5--8.5\,keV region. The vertical dashed line shows the energy of the neutral Fe\,K$\alpha$ line.}
\label{fig:ratiodm}
 \end{minipage}
\end{figure*}

\section{Time analysis}\label{sect:lc}

We analyzed the 2--10\,keV FI-XIS light-curve (in 5760\,s orbital bins) of IRAS\,00521$-$7054 (Fig.\,\ref{fig:xislc}), and found that the source shows a fractional rms variability of $F_{\rm\,Var}=16.7\pm1.1\%$. 

On a longer time-scale, we found no evidence of flux variability in the 2--10\,keV energy range between the {\it Suzaku} observations  ($F_{2-10}=2.3\times10^{-12}\rm\,erg\,cm^{-2}\,s^{-1}$), and the two {\it XMM-Newton} observation ($F_{2-10}=2.4\times10^{-12}\rm\,erg\,cm^{-2}\,s^{-1}$, \citealp{Tan:2012fk}) carried out six years earlier. The source also showed a consistent flux level between the two {\it XMM-Newton} observations carried out one month apart. No significant spectral variability is detected, and for the three observations the ratio between the 5--10\,keV and the 2--5\,keV flux is constant ($F_{5-10\rm\,keV}/F_{2-5\rm\,keV}\sim 1.6$). The luminosity of the source in the 2--10\,keV band is $2.5\times 10^{43}\rm\,erg\,s^{-1}$.

\section{Spectral analysis}\label{sect:specAnalysis}
The spectral analysis was carried out using XSPEC v12.7.1b \citep{Arnaud:1996kx}. We initially used only the higher signal-to-noise ratio {\it Suzaku} data to find the best models, and then fitted simultaneously the {\it Suzaku} and {\it XMM-Newton} spectra to better constrain the parameters of the models (Tables\,\ref{tab:resultsFit_abs} and \ref{tab:resultsFit_broadline}).

We used data in the 0.7--10\,keV and 0.6--8.0\,keV region, for {\it Suzaku} FI-XIS and BI-XIS, respectively, and for both FI-XIS and BI-XIS we excluded the interval 1.7--1.9\,keV, as indicated by the standard guidelines.
We added to the models a multiplicative factor to take into account the cross-calibration between FI-XIS and BI-XIS. We fixed the factor to one for FI-XIS, and let the BI-XIS factor free. For all the spectra the value of the factor turned out to be close to one within a few percent.
We added to all the fits Galactic absorption, fixing the column density to $N_{\rm\,H}^{\rm\,Gal}=4.44\times10^{20}\rm\,cm^{-2}$ \citep{Dickey:1990uq}.
We took into account in all models the redshift of IRAS\,00521$-$7054, and in the following we will always refer to the rest frame energy of the source.

As a first test, we fitted the combined FI-XIS and BI-XIS spectra with a power-law absorbed by neutral material. We modelled the neutral absorption using \texttt{TBABS} \citep{Wilms:2000vn}\footnote{in XSPEC \texttt{TBABS$_{\mathrm{Gal}}\times$(TBABS$\times$ZPOW)}}, and obtained $N_{\rm\,H}\simeq 6\times10^{22}\rm\,cm^{-2}$ and a chi-squared of $\chi^2=982.5$ for 753 degrees of freedom (DOF). In the left panel of Fig.\,\ref{fig:ratiodm} we show the residuals obtained by fitting the 2--4.5\,keV and 7.5--10\,keV spectrum with the absorbed power-law model. The figure clearly shows the presence of a soft excess below 2\,keV and of a large excess in the 4.5-7\,keV region (right panel).

More complex modelling is thus needed to reproduce the X-ray spectrum of IRAS\,00521$-$7054. In the following we describe the results obtained by applying partially covering neutral (Sect.\,\ref{sect:specan1}) or ionized (Sect.\,\ref{sect:specan2}) absorption models, and then a blurred reflection scenario (Sect.\,\ref{sect:specan3}).

\begin{figure*}[t!]
\centering
\centering
\includegraphics[width=\textwidth]{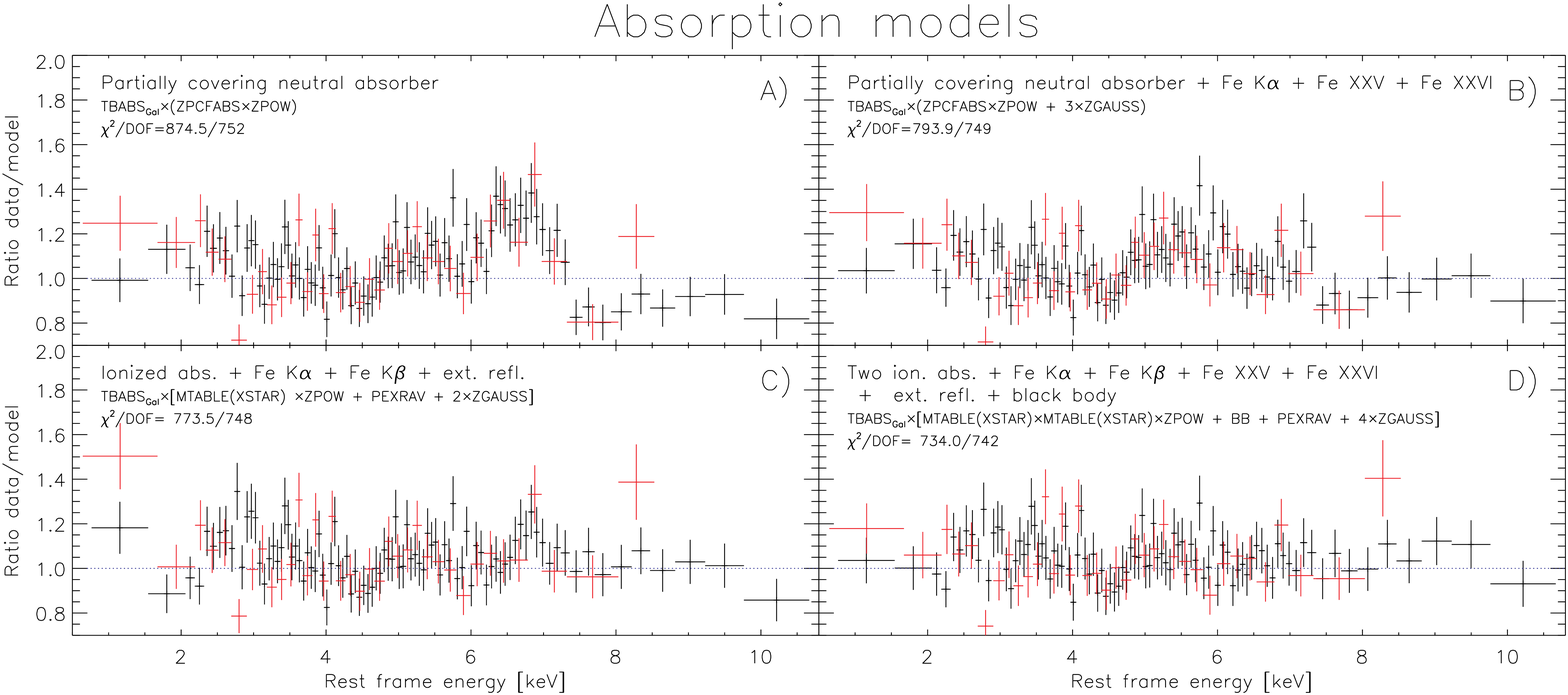}
\vspace{5mm}
\includegraphics[width=\textwidth]{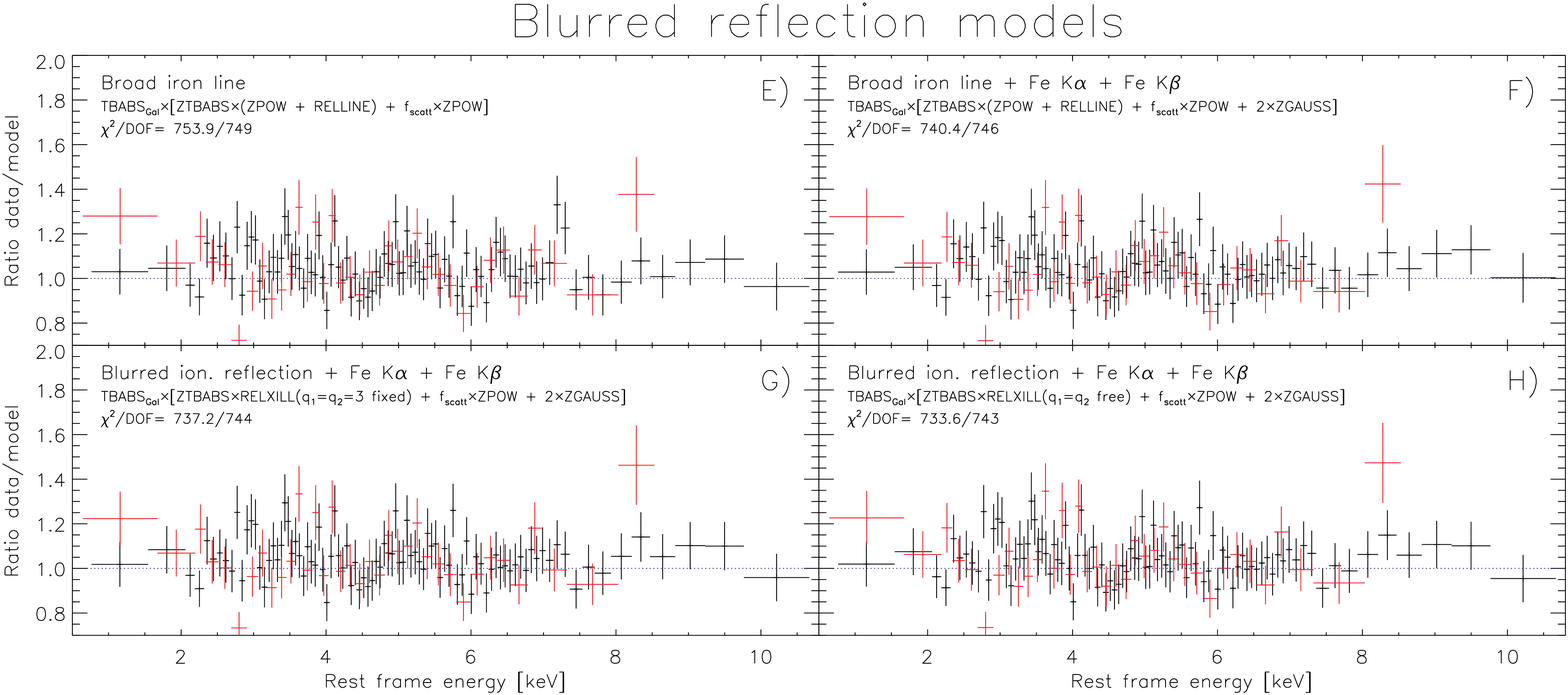}
  \caption{Ratio between the {\it Suzaku} data and the models discussed in Sect.\,\ref{sect:specAnalysis}. The black and red points represent the data obtained by FI-XIS and BI-XIS, respectively. {\it Top panel}: ratios obtained using neutral and ionized absorption models. Panels A and B show the residuals obtained applying a partially covering neutral absorber model (Sect.\,\ref{sect:specan1}), while panels C and D illustrate the ratios between the data and two models which consider fully covering ionized absorption (Sect.\,\ref{sect:specan2}). The spectral parameters obtained by applying the models used for Panels B and D to {\it XMM-Newton} and {\it Suzaku} data are reported in columns 2 and 3 of Table\,\ref{tab:resultsFit_abs}, respectively. 
{\it Bottom panel}: ratios obtained using blurred reflection models. Panels E and F show the ratios obtained by fitting the data with a relativistic broad Fe K$\alpha$ line (Sect.\,\ref{sect:specan3}), while Panels G and H those obtained assuming a blurred ionized reflection scenario. The spectral parameters obtained by applying the models used for Panels F and H to {\it XMM-Newton} and {\it Suzaku} spectra are reported in columns 2 and 3 of Table\,\ref{tab:resultsFit_broadline}, respectively. }
\label{fig:ratio_datamodels}
\end{figure*}

\subsection{Absorption models}\label{sect:specan_abs}

\subsubsection{Partially covering neutral absorber}\label{sect:specan1}
A scenario in which the X-ray source is partially covered by a neutral absorber was proposed by \cite{Tan:2012fk} as one of the possible explanations of the strong excess observed in the Fe K$\alpha$ region. Fitting the XIS spectra with a power-law continuum absorbed by a partially covering neutral absorber\footnote{\texttt{TBABS$_{\mathrm{Gal}}\times$(ZPCFABS$\times$ZPOW)}}, results in $\chi^2= 874.5$ for 752 DOF, and cannot account for the strong residuals in the iron line region (panel\,A of Fig.\,\ref{fig:ratio_datamodels}). Adding another fully covering neutral absorber, consistently with what was done by \cite{Tan:2012fk}\footnote{\texttt{TBABS$_{\mathrm{Gal}}\times$[ZTBABS$\times$(ZPCFABS$\times$ZPOW)]}}, does not improve the fit ($\Delta\chi^2\simeq0$).

Seyfert\,2s usually show evidence of a narrow ($\sigma \simeq 2,000\rm\,km\,s^{-1}$, \citealp{Shu:2011bh}) Fe K$\alpha$ line centred at 6.4\,keV (e.g., \citealp{Fukazawa:2011fk}). To take into account this feature, we added to the model a Gaussian line (\texttt{ZGAUSS}\footnote{\texttt{TBABS$_{\mathrm{Gal}}\times$(ZPCFABS$\times$ZPOW + ZGAUSS})}) with a width fixed to 1\,eV, well below the energy resolution of XIS, and with energy fixed at 6.4\,keV. This improves the fit ($\Delta\chi^2\simeq$36), but still leaves important residuals above 6.5\,keV.
Leaving both the energy and width of the line free to vary improves the fit ($\chi^2/\mathrm{DOF}=785.1/752$), but leaves significant residuals at $\sim 6$\,keV. Moreover the fit requires a very large line ($\sigma=797^{+189}_{-166}\rm\,eV$, $EW=660^{+130}_{-110}\rm\,eV$) centred at $E=6.18^{+0.15}_{-0.24}$\,keV, with a full-width at half maximum of $\mathrm{FWHM}\simeq 91,100\rm\,km\,s^{-1}$. The EW of the lines was calculated using the \texttt{eqwidth} command in XSPEC.
 
A significant fraction of Seyfert\,1s (e.g., \citealp{Patrick:2012qf}), and at least few Seyfert\,2s (e.g., \citealp{Lobban:2014zr}) show evidence of emission from ionized iron, typically Fe\,\textsc{xxv} and Fe\,\textsc{xxvi}.
Therefore we added three narrow lines (fixing $\sigma=\rm\,1eV$)\footnote{\texttt{TBABS$_{\mathrm{Gal}}\times$(ZPCFABS$\times$ZPOW + 3$\times$ZGAUSS})}\label{foot:neutrabs}, representing emission from neutral Fe\,K$\alpha$ and from Fe\,\textsc{xxv} and Fe\,\textsc{xxvi}, with energies fixed at 6.40\,keV, 6.70\,keV and 6.97\,keV, respectively. This improves significantly the fit ($\Delta \chi^2\simeq 80$ for three extra DOF), but leaves significant residuals between 4 and 6 keV (panel\,B).

\begin{table*}
\begin{center}
\caption[]{The table reports (1) the values of the parameters obtained by fitting  the combined {\it Suzaku} and {\it XMM-Newton} spectra of IRAS\,00521$-$7054 with: (2) a partially covering neutral absorber model (residuals for {\it Suzaku} data shown in panel\,B of Fig.\,\ref{fig:ratio_datamodels} and the model is reported in footnote\,11); (3) a double fully-covering ionized absorber model (panel\,D, footnote\,17).}
\label{tab:resultsFit_abs}
\begin{tabular}{lcc}
\noalign{\smallskip}
\multicolumn{3}{c}{\large Absorption models}\\
\noalign{\smallskip}
\hline \hline \noalign{\smallskip}
\multicolumn{1}{l}{\,\,\,\,\,\,\,\,\,\,\,\,\,\,\,\,\,\,\,\,\,\,\,\,\,(1)} & (2) &  (3)   \\
\noalign{\smallskip}
Model Parameters & Partially Covering Neutral absorber & Two Ionized absorbers \\
\noalign{\smallskip}
\hline \noalign{\smallskip}
\texttt{ZPOW}: Photon index $\Gamma$	 											& 	$1.98^{+0.03}_{-0.03}$		& 	  $2.22^{+0.10}_{-0.12}$	\\
\noalign{\smallskip}
\texttt{ZPCFABS}: Col. density $N_{\rm\,H}$ [{\tiny $10^{22}\rm\,cm^{-2}$}] 	 				& 	$7.3^{+0.2}_{-0.2}$		 	& --	\\
\noalign{\smallskip}
\texttt{ZPCFABS}: Cov. factor $f_{\rm\,C}$ 	 	 [{\tiny $\%$}] 							&  $99.2^{+0.1}_{-0.1}$	 	& --		\\
\noalign{\smallskip}
\texttt{MTABLE\{XSTAR(1)\}}: Col. density $N_{\rm\,H}^{\rm\,W,\,1}$ 	  [{\tiny $10^{22}\rm\,cm^{-2}$}]				& 	--	 		&  $27^{+5}_{-4}$		\\
\noalign{\smallskip}
\texttt{MTABLE\{XSTAR(1)\}}: Ioniz. par. $\log \xi^{\,1}$ 	 [{\tiny $\rm\,erg\,cm\,s^{-1}$}]				& 	--	 		&  $0.64^{+0.13}_{-0.19}$		\\
\noalign{\smallskip}
\texttt{MTABLE\{XSTAR(2)\}}: Col. density $N_{\rm\,H}^{\rm\,W,\,2}$ 	  [{\tiny $10^{22}\rm\,cm^{-2}$}]				& 	--	 	&  $34^{+69}_{-32}$	\\
\noalign{\smallskip}
\texttt{MTABLE\{XSTAR(2)\}}: Ioniz. par. $\log \xi^{\,2}$ 	 [{\tiny $\rm\,erg\,cm\,s^{-1}$}]				& 	--	 		&  $3.1^{+0.4}_{-0.6}$	\\
\noalign{\smallskip}
\texttt{BB}: Temperature $kT_{\rm\,e}$ 	 			& 	--		 	& 	 $0.10^{+0.05}_{-0.03}$			\\
\noalign{\smallskip}
\texttt{PEXRAV}: Reflection param. $R$ 	 			& 	--		 	& 	 $1.9^{+0.6}_{-1.2}$			\\
\noalign{\smallskip}
\texttt{ZGAUSS}: Energy(Fe\,K$\alpha$)	[{\tiny keV}] 			& 	$ 6.40^*$		 	& 	$6.40^*$		\\
\noalign{\smallskip}
\texttt{ZGAUSS}: EW(Fe\,K$\alpha$)	[{\tiny eV}] 			& 	$81^{+14}_{-14}$		 		& 	 $61^{+18}_{-14}$		\\
\noalign{\smallskip}
\texttt{ZGAUSS}: Energy(Fe\,K$\beta$)	[{\tiny keV}] 			& 	--		 	& 	 $7.24^{+0.10}_{-0.11}$	\\
\noalign{\smallskip}
\texttt{ZGAUSS}: EW(Fe\,K$\beta$)	[{\tiny eV}] 			& 	--		 		&  	 $34^{+20}_{-16}$		\\
\noalign{\smallskip}
\texttt{ZGAUSS}: Energy(Fe\,\textsc{xxv})	[{\tiny keV}] 				& 	$6.70^*$		 	&  $6.70^*$	\\
\noalign{\smallskip}
\texttt{ZGAUSS}: EW(Fe\,\textsc{xxv})	[{\tiny eV}] 				& 	$64^{+9}_{-11}$		 		& 	 $77^{+15}_{-15}$	\\
\noalign{\smallskip}
\texttt{ZGAUSS}: Energy(Fe\,\textsc{xxvi})	[{\tiny keV}] 			& 	$6.97^*$		 	& 	$6.97^*$	\\
\noalign{\smallskip}
\texttt{ZGAUSS}: EW(Fe\,\textsc{xxvi})	[{\tiny eV}] 			& 	$39^{+11}_{-13}$		 		& $25^{+19}_{-14}$	\\
\noalign{\smallskip}
\hline
\noalign{\smallskip}
$\chi^2/\mathrm{DOF}$	 							& 	1168.5/1052		 		& 1104.1/1045	\\
\noalign{\smallskip}
\hline
\noalign{\smallskip}
\multicolumn{3}{l}{{\bf Notes}. $^*$ value of the parameter fixed.}\\
\end{tabular}
\end{center}
\end{table*}

To include reflection from neutral material, we used the \texttt{PEXRAV} model \citep{Magdziarz:1995pi}, fixing the cut-off energy to an arbitrary value ($E_{\rm\,C}=300$\,keV) and the inclination angle to $i=30^{\circ}$. Leaving $i$ free to vary does not improve the fit. We fixed the values of the photon index and normalisation to those of the X-ray continuum, and considered only the reflection component, leaving the reflection parameter ($R=\Omega/2\pi$, where $\Omega$ is the solid angle of the reflector) free to vary.
We first tested the scenario in which the reflector is outside the line of sight of the clumpy neutral absorber ({\it external reflection\footnote{\texttt{TBABS$_{\mathrm{Gal}}\times$(ZPCFABS$\times$ZPOW + PEXRAV +3$\times$ZGAUSS)}}}), as it would be expected if most of the reprocessed X-ray radiation is produced in the molecular torus. This improved the fit ($\Delta \chi^2\simeq 16$ for one extra DOF), although it left residuals below 6\,keV and it resulted in a value of the reflection parameter of $R=3.8^{+0.6}_{-0.5}$. This appears to be not consistent with the EW of the narrow Fe K$\alpha$ line ($\sim 60$\,eV). To compare the Fe K$\alpha$ EW with the value of the reflection parameter we computed the EW of the line with respect to the reflection component (EW$^{\rm\,refl}$). We obtained EW$^{\rm\,refl}\simeq160$\,eV, a value much lower than that expected theoretically (1--2\,keV, \citealp{Matt:1991uq}).

We then tested the scenario in which the reflected emission is also seen through the neutral patchy absorber ({\it internal reflection\footnote{\texttt{TBABS$_{\mathrm{Gal}}\times$(ZPCFABS$\times$PEXRAV+3$\times$ZGAUSS)}}}), as it would be expected if the absorption is due to the host galaxy, or if the bulk of the reprocessed emission comes from the outer part of the accretion disk. In this case we obtained a better value of the reduced chi-squared ($\chi^2/\mathrm{DOF}=783.5/748$), although the fit leaves significant residuals at $E\simeq 5$\,keV. Moreover, the fit results in a very large value of the reflection component ($R=4.9_{-0.4}^{+0.9}$), which is inconsistent with the EW of the narrow Fe K$\alpha$ line. 

To better constrain the parameters, we fitted simultaneously the spectra obtained by {\it Suzaku} (FI-XIS, BI-XIS) and the two {\it XMM-Newton} (PN, MOS1 and MOS2) observations with the model reported in Panel\,B of Fig.\,\ref{fig:ratio_datamodels} (partially covering neutral absorber plus three narrow emission lines). We added multiplicative factors to take into account the cross-calibration between the different detectors, and for possible flux variability between the different observations. For all the spectra the value of the factor turned out to be close to one within a few percent, as expected from the lack of long-term variability discussed in Sect.\,\ref{sect:lc}. The values of the parameters obtained are listed in column\,2 of Table\,\ref{tab:resultsFit_abs}.  In all the fits described above we obtained values of the covering factor of the neutral absorber of $95-99\%$. This makes this model indistinguishable from one including an absorbed power-law plus a component due to scattered emission, since the amplitude of the scattered emission is usually of a few percents (see Sect.\,\ref{Sect:ionizedReflection}).

Based on the analysis reported in this section, we can therefore exclude that a partially covering neutral absorber model can reproduce the broad feature observed in the X-ray spectrum of IRAS\,00521$-$7054.

\subsubsection{Ionized absorption}\label{sect:specan2}

Ionized absorption could distort the continuum in such a way that it would mimic the shape of an intrinsically broad Fe\,K$\alpha$ line (e.g., \citealp{Turner:2009fk} and references therein). \citeauthor{Miyakawa:2012vn} (\citeyear{Miyakawa:2012vn}, see also \citealp{Inoue:2003uq}, \citealp{Miyakawa:2009fu} and \citealp{Inoue:2011dz}) recently proposed, to explain the broad line observed in the X-ray spectrum MCG-6-30-15,  a model consisting of a power-law plus reflection from neutral material, obscured by three different absorbers, one fully covering and two partially covering the X-ray source, with different values of the ionization parameter and of the column density. In their model X-ray reflection was observed as the superposition of a Compton hump and two narrow Gaussian lines, representing Fe K$\alpha$ and Fe K$\beta$ fluorescent emission.

We tested whether ionized absorption might explain the broad feature observed in the X-ray spectrum of IRAS\,00521$-$7054. To take into account ionized absorption, we used the same grid of XSTAR \citep{Kallman:2001fk,Bautista:2001uq} photoionized absorption models adopted by \cite{Miyakawa:2012vn}. This XSTAR table model assumes a photon index of the primary X-ray continuum of $\Gamma=2$, and that the warm absorber has a solar abundance, a temperature of $10^5$\,K, a pressure of $0.03\rm\,dyn\,cm^{-2}$ and a density of $10^{12}\rm\,cm^{-3}$. The choice of $\Gamma$ does not affect the results because of the weak dependence of the XSTAR model on this parameter. The free parameters of the absorber are the column density of the ionized material ($N_{\rm\,H}^{\rm\,W}$) and its ionization parameter ($\xi$). For all the models we obtained a value of the covering factor consistent with unity, and we therefore discarded the hypothesis that the ionized medium partially covers the X-ray source.

\begin{figure}[t!]
\centering
\centering
\includegraphics[width=0.5\textwidth]{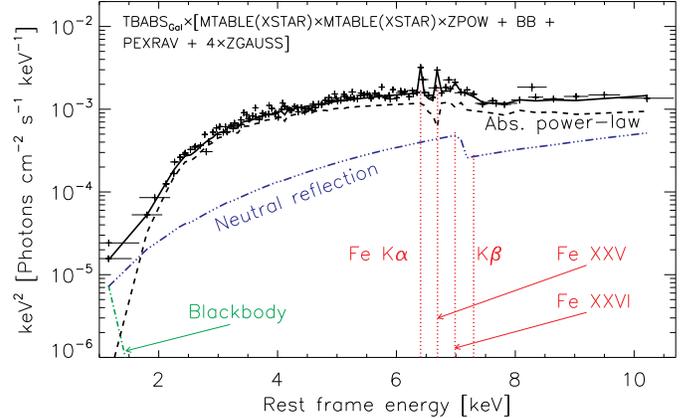}
  \caption{{\it Suzaku} spectrum of IRAS 00521$-$7054 together with the ionized absorption model (power-law continuum seen through two layers of ionized material, plus neutral reflection, a blackbody and four narrow lines representing Fe\,K$\alpha$, Fe\,K$\beta$, Fe \textsc{xxv} and Fe\,\textsc{xxvi} lines). The ratio between the data and the model associated to this figure is shown in panel\,D of Fig.\,\ref{fig:ratio_datamodels}. }
\label{fig:eeufs_abs}
\end{figure}

We started by taking into account one layer of ionized absorption, including both reflection from neutral material and two narrow lines (Fe K$\alpha$ and Fe\,K$\beta)$\footnote{\texttt{TBABS$_{\mathrm{Gal}}\times$[MTABLE(XSTAR)$\times$ZPO + PEXRAV + 2$\times$ZGAUSS]}}. The fit yields $\chi^2/\mathrm{DOF}=773.5/748$, and leaves significant residuals between 5 and 6\,keV and around 7\,keV (Panel\,C of Fig.\,\ref{fig:ratio_datamodels}). 
Adding to the model two narrow Gaussian lines with energies fixed to 6.70 and 6.97\,keV, representing Fe\,\textsc{xxv} and Fe\,\textsc{xxvi}, respectively\footnote{\texttt{TBABS$_{\mathrm{Gal}}\times$[MTABLE(XSTAR)$\times$ZPO + PEXRAV + 4$\times$ZGAUSS]}}, improves the fit ($\Delta \chi^2\simeq 12$), and reduces the residuals in the iron line region. While the presence of Fe\,\textsc{xxv} and Fe\,\textsc{xxvi} emission lines improves significantly the fit, it must be stressed that some degeneracy exists between these lines and the absorption features of the ionized absorber, which have the same energy.
A second layer of ionized material\footnote{\texttt{TBABS$_{\mathrm{Gal}}\times$[MTABLE(XSTAR)$\times$MTABLE(XSTAR)$\times$ZPO + PEXRAV + 4$\times$ZGAUSS]}} improves the fit ($\Delta \chi^2\simeq 6$), but the model still leaves significant residuals below 2\,keV. These residuals can be taken into account by adding a blackbody component\footnote{\texttt{TBABS$_{\mathrm{Gal}}\times$[MTABLE(XSTAR)$\times$MTABLE(XSTAR)$\times$ZPO + BB + PEXRAV + 4$\times$ZGAUSS]}}, which produces the best fit ($\chi^2=734.0$ for 742\,DOF, Panel\,D) to the data using absorption models. This model, together with the {\it Suzaku} data, is shown in Fig.\,\ref{fig:eeufs_abs}. 
Fitting simultaneously the spectra obtained by {\it Suzaku} and the two {\it XMM-Newton} observations with the two ionized absorbers model we obtained that the fit yields a good value of the reduced-chi squared ($\chi^2$/DOF=1104.1/1045). The values of the parameters obtained are listed in column\,3 of Table\,\ref{tab:resultsFit_abs}.

The ionization stage of the material producing the Fe K$\beta$ line can be deduced from the energy centroid of the line ($E_{\rm\,K\beta}$, e.g., \citealp{Yaqoob:2007ys,Palmeri:2003zr}), bearing in mind that iron must be less ionized than Fe\,\textsc{xvii} to produce K$\beta$ lines, because Fe\,\textsc{xvii} and ions with higher ionization states lack M shell electrons. Comparing the value of $E_{\rm\,K\beta}$ with theoretical predictions (see e.g. Fig.\,2 of \citealp{Liu:2010vn}) we found that the Fe K$\beta$ line is consistent with being emitted by Fe\,\textsc{xiv}--Fe\,\textsc{xvi}. 
The energy of the Fe K$\alpha$ line ($E_{\rm\,K\alpha}$) is less sensitive to the ionization stage of the iron. Leaving $E_{\rm\,K\alpha}$ free to vary does not significantly improve the fit ($\Delta \chi^2\simeq 1$), and results in $E_{\rm\,K\alpha}=6.43^{+0.04}_{-0.06}$\,keV, which is also consistent with being produced by Fe\,\textsc{xiv}--Fe\,\textsc{xvi} as well as by neutral iron.
The ratio between the intensity of the Fe\,K$\beta$ and Fe\,K$\alpha$ ($I_{\rm\,K\beta}/I_{\rm\,K\alpha}$) line is expected to vary between 0.12 and 0.20, depending on the ionization stage of the iron \citep{Palmeri:2003zr,Mendoza:2004ly}. Applying the best ionized absorption model we obtained that $I_{\rm\,K\beta}/I_{\rm\,K\alpha}$ ranges between 0.03 and 0.50, which is in agreement with the theoretical predictions, but cannot be used to constrain the ionization stage.

\begin{table*}
\begin{center}
\caption[]{The table reports (1) the values of the parameters obtained by fitting the combined {\it Suzaku} and {\it XMM-Newton} spectra with: (2) a model which considers an absorbed power-law continuum and a broad relativistic Fe K$\alpha$ line (residuals for {\it Suzaku} data shown in panel\,F of Fig.\,\ref{fig:ratio_datamodels}, model reported in footnote\,20); (3) an absorbed blurred ionized reflection model (panel\,H, footnote\,21).}
\label{tab:resultsFit_broadline}
\begin{tabular}{lcc}
\noalign{\smallskip}
\multicolumn{3}{c}{\large Blurred reflection models}\\
\noalign{\smallskip}
\hline \hline \noalign{\smallskip}
\multicolumn{1}{l}{\,\,\,\,\,\,\,\,\,\,\,\,\,\,\,\,\,\,\,\,\,\,\,\,\,(1)} & (2) &  (3)   \\
\noalign{\smallskip}
Model Parameters & Broad line &Blurred ionized reflection\\
\noalign{\smallskip}
\hline \noalign{\smallskip}
\texttt{ZPOW}: Photon index $\Gamma$	 						&  $1.99^{+0.07}_{-0.08}$	& --	\\
\noalign{\smallskip}
\texttt{CONST}: $f_{\rm\,scatt}$ 		 [{\tiny $\%$}] 					& $0.8^{+0.1}_{-0.1}$		& 	$0.7^{+0.1}_{-0.1}$		\\
\noalign{\smallskip}
\texttt{ZTBABS}: Col. density $N_{\rm\,H}$ [{\tiny $10^{22}\rm\,cm^{-2}$}] 	 	&  $6.9^{+0.3}_{-0.3}$ & 	$7.2^{+0.3}_{-0.3}$		\\
\noalign{\smallskip}
\texttt{RELLINE}:  Energy	 [{\tiny keV}]				 		& $6.40^*$	& 	--		\\
\noalign{\smallskip}
\texttt{RELLINE}:  Index	$q_1=q_2$			 		& $3^*$	& 	--		\\
\noalign{\smallskip}
\texttt{RELLINE}:  Spin	$a$			 		& $ \geq0.78$	& 	--		\\
\noalign{\smallskip}
\texttt{RELLINE}:   Inclination angle $i$ [{\tiny deg}]	 				 		& 	$41^{+2}_{-2}$ & 	--		\\
\noalign{\smallskip}
\texttt{RELLINE}:    EW	[{\tiny eV}]			 		& 	$798^{+152}_{-102}$ & 	--		\\
\noalign{\smallskip}
\texttt{RELXILL}: Photon index $\Gamma$	 			 	& -- 	&  $2.28^{+0.25}_{-0.11}$	\\
\noalign{\smallskip}
\texttt{RELXILL}:   Index $q_1=q_2$	 				 		& --	& 	$3.1^{+0.5}_{-0.4}$		\\
\noalign{\smallskip}
\texttt{RELXILL}:  Spin $a$ 				 		& --	& 	$\geq0.73$		\\
\noalign{\smallskip}
\texttt{RELXILL}:   Inclination angle $i$ [{\tiny deg}]	 				 		& 	--  & 	$ 40^{+2}_{-3}$		\\
\noalign{\smallskip}
\texttt{RELXILL}: Iron abundance $Z_{\rm\,Fe}$ [{\tiny Fe/solar}]	  				 		& --	& 	$ 2.6^{+2.1}_{-1.0}$		\\
\noalign{\smallskip}
\texttt{RELXILL}: Ioniz. par. $\log \xi$ 	 [{\tiny $\rm\,erg\,cm\,s^{-1}$}]  	 				 		& --	& 	$ 2.0^{+0.4}_{-2.0}$		\\
\noalign{\smallskip}
\texttt{RELXILL}: Reflection par.	 			 	& --	& 	$  2.7^{+2.5}_{-1.3}$		\\
\noalign{\smallskip}
\texttt{ZGAUSS}: Energy(Fe\,K$\alpha$)	[{\tiny keV}] 			 	& 	$6.40^*$	& 	$ 6.40^*$		\\
\noalign{\smallskip}
\texttt{ZGAUSS}: EW(Fe\,K$\alpha$)	[{\tiny eV}] 			 	& 	$25^{+11 }_{-14 }$	& 	$ 28^{+13 }_{-21 }$		\\
\noalign{\smallskip}
\texttt{ZGAUSS}: Energy(Fe\,K$\beta$)	[{\tiny keV}] 			 		& 	$7.21^{+0.17}_{-0.33}$& 	$ 7.20^{+0.09}_{-0.09}$		\\
\noalign{\smallskip}
\texttt{ZGAUSS}: EW(Fe\,K$\beta$)	[{\tiny eV}] 			 	&  	$ 28^{+ 18}_{-17 }$	& 	$\leq33$		\\
\noalign{\smallskip}
\hline
\noalign{\smallskip}
$\chi^2/\mathrm{DOF}$	 								&  1105.3/1049	& 	1100.5/1046		\\
\noalign{\smallskip}
\hline
\noalign{\smallskip}
\multicolumn{3}{l}{{\bf Notes}. $^*$ value of the parameter fixed.}\\
\end{tabular}
\end{center}
\end{table*}

\subsection{Blurred reflection}\label{sect:specan3}
Ionized reflection from the accretion disk, blurred by relativistic effects in the proximity of the SMBH, could be the mechanism responsible for the broad iron lines (e.g., \citealp{Reynolds:2003cr}). Applying physical models of blurred reflection, broad iron lines might be used to infer the spin of the SMBH (e.g, \citealp{Dovciak:2004kl}, \citealp{Brenneman:2006dq}, \citealp{Reynolds:2008nx}, \citealp{Brenneman:2011oq}, \citealp{Patrick:2012qf}, \citealp{Reynolds:2013uq}), which is of crucial importance to constrain the radiative efficiency of accretion (e.g., \citealp{Trakhtenbrot:2014fk}).

In the following we will first discuss the results obtained by adopting a model which considers a broad relativistic iron line (Sect.\,\ref{sect:broadline}), and then we will apply a more physical scenario, adding to the absorbed power-law emission a blurred ionized reflection component from the accretion flow (Sect.\,\ref{Sect:ionizedReflection}).

\subsubsection{Broad line}\label{sect:broadline}
As a first test, we added to the absorbed power-law continuum a broad relativistic line using the \texttt{RELLINE} model \citep{Dauser:2010ys,Dauser:2013kl}, fixing the energy of the line to 6.4\,keV.
The model assumes a reflection emissivity of the type $\epsilon(r)=r^{-q_1}$ for $r<R_{\rm\,break}$, and $\epsilon(r)=r^{-q_2}$ for  $r>R_{\rm\,break}$, where $r$ is the distance from the SMBH.
The free parameters of \texttt{RELLINE} are the inclination angle of the observer with respect to the accretion disk ($i$), the spin of the SMBH ($a$), the inner ($q_1$) and outer ($q_2$) emissivity index of the disk, the radius at which the emissivity index changes ($R_{\rm\,break}$), and the inner ($R_{\rm\,in}$) and outer ($R_{\rm\,out}$) radius of the disk. The model allows different profiles of the angular emission $M(\mu_{e})$, where $\mu_{\rm\,e}=\cos \theta_{\rm\,e}$, and $\theta_{\rm\,e}$ is the emission angle (see \citealp{Svoboda:2009qf}). In particular, in \texttt{RELLINE} it is possible to adopt a limb-darkening law ($M(\mu_{e})= 1+2.06\mu_{\rm\,e}$, \citealp{Laor:1991ij}), an isotropic emission ($M(\mu_{e})= 1$) or a limb-brightening law ($M(\mu_{e})= \ln(1+\mu_{\rm\,e})$, \citealp{Haardt:1993bh}).
In the following, unless otherwise stated, we will adopt the limb-darkening law for consistency with the other relativistic iron line models (e.g., \citealp{Laor:1991ij}, \citealp{Brenneman:2006dq}). 
In \texttt{RELLINE} the radius of the inner stable circular orbit ($R_{\rm\,ISCO}$) scales with the spin of the SMBH. We assumed that $R_{\rm\,in}=R_{\rm\,ISCO}$, $R_{\rm\,out}=400\,r_{\rm\,G}$ (where $r_{\rm\,G}$ is the gravitational radius of the SMBH), and took into account the soft excess using a scattered component\footnote{\texttt{TBABS$_{\mathrm{Gal}}\times$[$f_{\rm\,scatt}\times$ZPOW + ZTBABS$\times$(ZPOW + RELLINE)]}} (see Sect.\,\ref{Sect:ionizedReflection} for details), which yields a better fit than a blackbody component ($\Delta \chi^2\simeq 2$).  For a flat disk irradiated by a central point source one would expect $\epsilon(r)\propto r^{-3}$. Therefore we applied \texttt{RELLINE} fixing $R_{\rm\,break}$ to an arbitrary value, and the emissivity indices to $q_1=q_2=3$. The fit yields $\chi^2$/DOF=753.9/749, and shows residuals in the Fe K$\alpha$ region (Panel\,E of Fig.\,\ref{fig:ratio_datamodels}). Adding a Gaussian line\footnote{\texttt{TBABS$_{\mathrm{Gal}}\times$[$f_{\rm\,scatt}\times$ZPOW + ZTBABS$\times$(ZPOW + RELLINE) + ZGAUSS]}}, fixing the energy to 6.4\,keV and the width to $\sigma=1\rm\,eV$ improves significantly the fit ($\Delta \chi^2=4.5$), however a visual inspection of the residuals still shows that the model underestimates the data in the 7--8\,keV region. This could be explained by the presence of a narrow Fe K$\beta$ line, associated to the narrow Fe K$\alpha$. 
Adding a second Gaussian line\footnote{\texttt{TBABS$_{\mathrm{Gal}}\times$[$f_{\rm\,scatt}\times$ZPOW + ZTBABS$\times$(ZPOW + RELLINE) + 2$\times$ZGAUSS]}}, with the width fixed to 1\,eV and the energy left free to vary, improves significantly the fit ($\Delta \chi^2=9$), and removes any residual in the 6--8\,keV region (panel\,F). However, the ratio between the intensity of the two lines is $I_{\rm\,K\beta}/I_{\rm\,K\alpha}\gtrsim 0.33$, larger than the maximum possible value (see Sect.\,\ref{sect:specan2}). This might be related to the drop of the broad line flux above $\sim 7.1$\,keV. The parameters obtained by fitting {\it Suzaku} and {\it XMM-Newton data} with this model are reported in column\,2 of Table\,\ref{tab:resultsFit_broadline}.

Assuming a limb-brightening or an isotropic emission of the disk ($\Delta \chi^2=0.4$ and $\Delta \chi^2=-0.1$, respectively), or relaxing the assumption on the emissivity indices, and letting them free to vary (assuming $q_1=q_2$; $\Delta \chi^2=0.6$), does not significantly improve  the fit, and results in values of the parameters consistent with those reported above.

\subsubsection{Ionized reflection}\label{Sect:ionizedReflection}

To self-consistently reproduce all the reflection features arising in an ionized disk, and blurred by the proximity of the SMBH, we used the recently developed \texttt{RELXILL} model \citep{Garcia:2014tg,Dauser:2013kl}. 
 \texttt{RELXILL} is created by the convolution of the \texttt{XILLVER} ionized reflection model \citep{Garcia:2010hc,Garcia:2011ij,Garcia:2013bs} with \texttt{RELCONV}, the convolution kernel associated to \texttt{RELLINE} \citep{Dauser:2010ys}. While previous ionized reflection models, such as \texttt{REFLIONX} \citep{Ross:2005hc}, provided angle-averaged solutions for the reprocessed emission, \texttt{XILLVER} calculates the emission angle for each point of the accretion disk. The simulations carried out to create \texttt{XILLVER} have shown that the profile of the angular emission depends on the ionization state of the disk, and follows the limb-brightening law, although not in the form usually taken into account (see Sect.\,\ref{sect:broadline}).
 \cite{Garcia:2014tg} have shown that, while the values of the spin and of the inclination angle obtained are similar using  \texttt{KDBLUR(REFLIONX)} and \texttt{RELXILL}, the value of the iron abundance ($Z_{\rm\,Fe}$) was overestimated by up to a factor of two when computed using \texttt{KDBLUR(REFLIONX)}.

\begin{figure}[t!]
\centering
\centering
\includegraphics[width=0.5\textwidth]{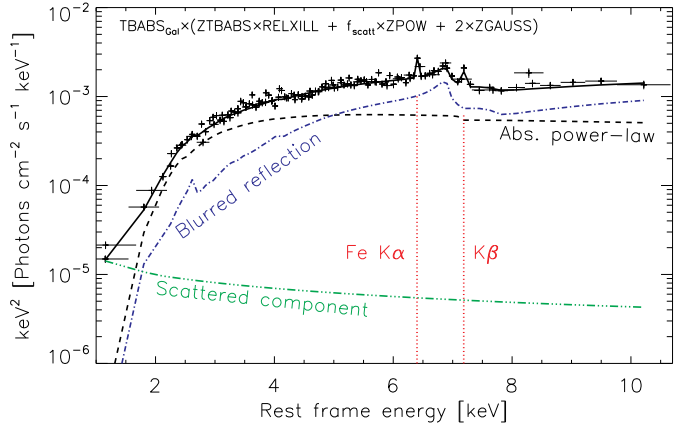}
  \caption{{\it Suzaku} spectrum of IRAS 00521$-$7054 fitted with the blurred reflection model (absorbed power-law emission plus absorbed blurred reflection, two narrow Fe K$\alpha$ and Fe K$\beta$ lines, and a scattered component). The ratio between the data and the model associated to this figure is shown in panel\,H of Fig.\,\ref{fig:ratio_datamodels}. }
\label{fig:eeufs}
\end{figure}

The free parameters of \texttt{RELXILL} are $i$, $Z_{\rm\,Fe}$, $\Gamma$, the ionization parameter of the disk ($\xi$), the fraction of reflected flux ($R$, defined as for \texttt{PEXRAV}), $R_{\rm\,in}$, $R_{\rm\,out}$, $q_1$, $q_2$, $R_{\rm\,break}$, and the cutoff energy ($E_{\rm\,C}$). As our spectra are limited to $E\lesssim10$\,keV we assumed $E_{\rm\,C}=300$\,keV. 
Following what was done in Sect.\,\ref{sect:broadline} we assumed that the inner radius is the inner stable circular orbit (i.e. $R_{\rm\,in}=R_{\rm\,ISCO}$), $R_{\rm\,out}=400\,r_{\rm\,G}$, we added to the model two narrow Gaussian lines to take into account the Fe K$\alpha$ and Fe K$\beta$ emission, and we used a scattered component for the soft-excess\footnote{\texttt{TBABS$_{\mathrm{Gal}}\times$($f_{\rm\,scatt}\times$ZPOW + ZTBABS$\times$RELXILL + 2$\times$ZGAUSS)}}. The scattered flux was taken into account using a power-law emission multiplied by a constant ($f_{\rm\,scatt}$), with both photon index and normalisation fixed to the values of the primary power-law. The scattered emission is believed to arise in circumnuclear material, and represents a fraction of typically few percent of the primary X-ray emission (e.g., \citealp{Cappi:2006qa}), although in some cases it has been found to be below 1\,\% \citep{Ueda:2007vn,Eguchi:2009kx}.

Fitting the broad-band {\it Suzaku} spectrum, setting the emissivity index of the accretion flow to $q_1=q_2=3$, we obtained $\chi^2/\mathrm{DOF}=737.2/744$ (panel\,G of Fig.\,\ref{fig:ratio_datamodels}). Relaxing the assumption on $q_1=q_2$ and leaving the parameter free to vary improves significantly the fit ($\Delta \chi^2=3.6$, panel\,H). This model can reproduce very well the spectrum in the 5--8\,keV region, and provides the best fit with reflection models to the {\it Suzaku} spectrum of IRAS\,00521$-$7054. The presence of the Fe\,K$\beta$ line is marginally significant ($\Delta \chi^2\simeq 5$ for two extra DOF), and we obtained only an upper limit on its EW. Similarly to what we found by applying the ionized absorption model (see Sect.\,\ref{sect:specan2}), we obtained that the energy of the Fe K$\beta$ line and the ratio between the intensity of the Fe K$\beta$ and K$\alpha$ lines ($I_{\rm\,K\beta}/I_{\rm\,K\alpha}\simeq 0.10-0.78$) are consistent with the line being emitted by ionized iron, although in this case we found a larger range of ionization states (Fe\,\textsc{xii}--Fe\,\textsc{xvi}). In Fig.\,\ref{fig:eeufs} we illustrate the {\it Suzaku} spectrum of IRAS\,00521$-$7054 together with the different spectral components obtained by the best fit.

\begin{figure}[t!]
\centering
\centering
\includegraphics[width=0.5\textwidth]{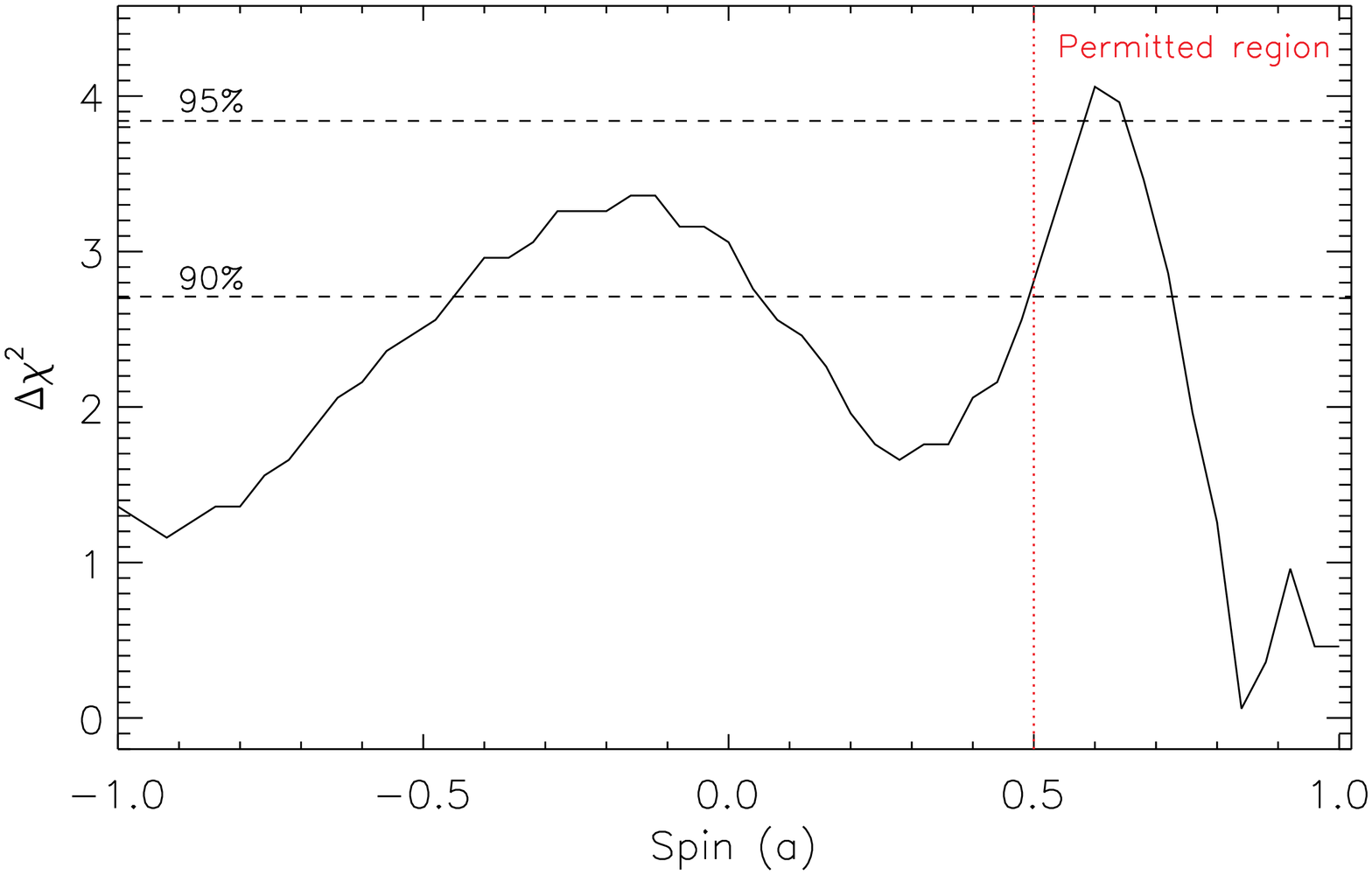}
  \caption{$\Delta \chi^2$ for different values of the spin of the SMBH ($a$) of IRAS\,00521$-$7054. The values of $a$ span from the maximally counter-rotating ($a=-1$) to the maximally rotating ($a=1$) scenario. The $\Delta \chi^2$ distribution was obtained by fitting the combined {\it Suzaku} and {\it XMM-Newton} spectra with \texttt{RELXILL} (Sect.\,\ref{Sect:ionizedReflection} and column 3 of Table\,\ref{tab:resultsFit_broadline}). As discussed in Sect.\,\ref{Sect:ionizedReflection}, by considering that $R=2.7^{+2.5}_{-1.3}$ and using the recent theoretical results of Dauser et al. (2014), we can exclude all the solutions for which $a\leq0.5$ (red vertical dotted line), which implies that the only possible solution is  $a \geq 0.73$.}
\label{fig:chisqspin}
\end{figure}

\begin{figure*}[t!]
\centering
\centering
\includegraphics[width=\textwidth]{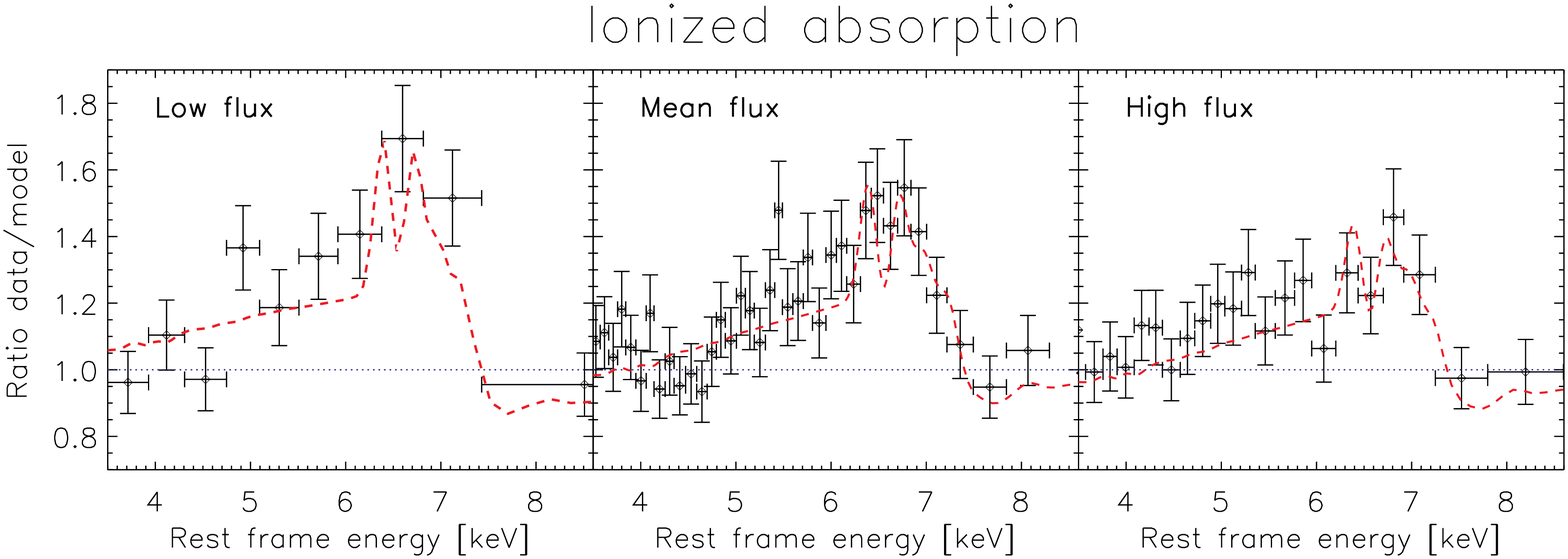}
\vspace{5mm}
\includegraphics[width=\textwidth]{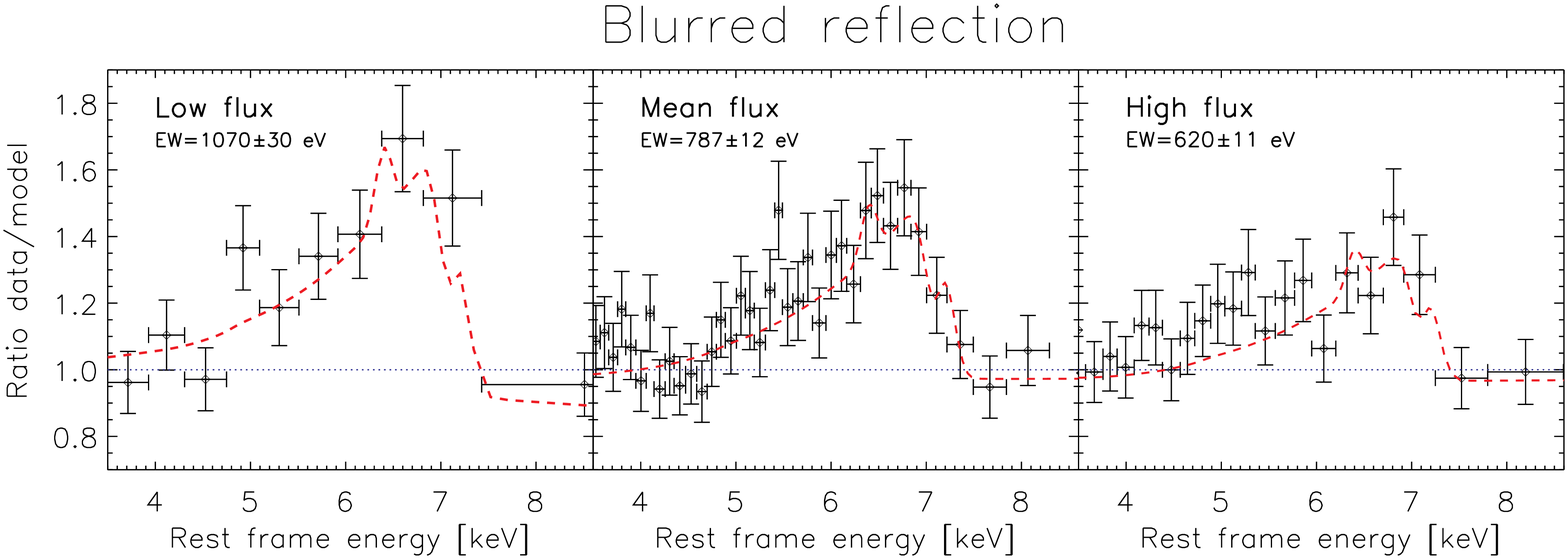}
  \caption{{\it Top panel:} ratio between {\it Suzaku} FI-XIS data and the fit obtained by applying a simple absorbed power-law model (\texttt{TBABS$_{\mathrm{Gal}}\times$ZTBABS$\times$ZPOW}) to the data in the 2--4.5\,keV and 7.5--10\,keV range. The three spectra represent different flux levels of the source (see Fig.\,\ref{fig:xislc}). The red dashed line represents the ratio between the best model obtained by fitting the spectra with a power-law absorbed by two layers of ionized material plus neutral reflection and four Gaussian lines (see Panel\,D of Fig.\,\ref{fig:ratio_datamodels}), freezing all the parameters to the values obtained by fitting the average spectrum except the power-law normalisation, and the absorbed power-law model.
{\it Bottom panel:} same as top panel for the blurred reflection scenario. The red dashed line represents the ratio between the best model obtained by fitting the spectra with a broad line model (see Panel\,H of Fig.\,\ref{fig:ratio_datamodels}), freezing all the parameters to the values obtained by fitting the average spectrum, leaving only the power-law normalisation free to vary, and the absorbed power-law model.}
\label{fig:ratiofluxstates}
\end{figure*}

Adding reflection from neutral material\footnote{\texttt{TBABS$_{\mathrm{Gal}}\times$($f_{\rm\,scatt}\times$ZPOW + ZTBABS$\times$RELXILL + 2$\times$ZGAUSS + PEXRAV)}} does not improve the chi-squared ($\Delta \chi^2\simeq 0$) and results in a reflection parameter of $R<2.1$. Leaving the energy of the Fe K$\alpha$ line free to vary does not affect the chi-squared ($\Delta \chi^2 \simeq 0.1$), and results in $E_{\rm\,K\alpha}=6.39^{+0.06}_{-0.06}$\,keV.
To self-consistently reproduce the Compton hump, the Fe K$\alpha$ and Fe K$\beta$ line produced in neutral material we used the \texttt{pexmon} model \citep{Nandra:2007ly}\footnote{\texttt{TBABS$_{\mathrm{Gal}}\times$($f_{\rm\,scatt}\times$ZPOW + ZTBABS$\times$RELXILL + PEXMON)}}, which assumes a slab geometry for the reprocessing material. This model yields a worse fit than that obtained by combining \texttt{pexrav} with two Gaussian lines ($\chi^{2}=738.8$ for 745 DOF), and results in a value of the reflection parameter of $R=0.6^{+0.3}_{-0.4}$.

To better constrain the parameters, we simultaneously fitted, as done in Sect.\,\ref{sect:specan2}, the {\it Suzaku} and the two {\it XMM-Newton} observations with our best blurred ionized reflection model. For all the spectra we obtained values of the multiplicative factors close to unity within a few percents. The values of the parameters obtained are listed in column\,3 of Table\,\ref{tab:resultsFit_broadline}. The fact that we only obtained an upper limit on the ionization parameter of the disk is caused by the lack of photons below 2\,keV, which are strongly absorbed by the obscuring material in the line of sight. The soft X-ray band gives in fact important constraints on $\xi$, thanks to the large number of atomic transitions.

From our combined fit we obtained that the value of the reflection parameter associated to the ionized disk exceeds unity ($R=2.7^{+2.5}_{-1.3}$), which might imply that light-bending is at play (e.g., \citealp{Martocchia:1996kx,Miniutti:2004uq,Gandhi:2007fk}). In the light-bending scenario the X-ray source is located very close to the SMBH, so that part of the power-law emission is bent towards the disk, which diminishes the observed primary emission and increases the fraction of reprocessed radiation. 
In Fig.\,\ref{fig:chisqspin} we illustrate the value of $\Delta \chi^2$ for different values of the spin. We found three possible solutions for the spin of the SMBH: $a\leq -0.5$, $a=0.3\pm0.2$ and $a\geq 0.73$ (Fig.\ref{fig:chisqspin}). Recently, Dauser et al. (2014) calculated the maximum possible value of the reflection parameter $R$ for a certain value of the spin in the lamp post geometry. Considering that the reflection parameter of IRAS\,00521-7054 is in the range $1.4 \leq R \leq 5.2$, and keeping the assumption $R_{\rm\,in}=R_{\rm\,ISCO}$, from Fig.\,3 (left panel) of Dauser et al. (2014) we can deduce that the spin of the SMBH must be $a\geq0.50$. This excludes the retrograde and intermediate-spin solutions, and implies that $a \geq 0.73$.

\section{Time-resolved spectral analysis}\label{sect:timeresolved}

As discussed in Sect.\,\ref{sect:lc}, IRAS\,00521$-$7054 is rather variable in the 2--10\,keV band on a time-scale of few ks. To study the relationship between flux and spectral variations, we performed time-resolved spectral analysis, dividing the light-curve into three different states: high, mean and low (see Fig.\,\ref{fig:xislc}). For each flux state we created FI-XIS and BI-XIS spectra, modifying the good time intervals to take into account only one flux state at the time, and adopting the same procedure for the data reduction described in Sect.\,\ref{sect:suzakudatared}.

As a first test we fitted each of the three spectra in the 2--4.5\,keV and 7.5--10\,keV range (in the rest frame of the source) using a simple absorbed power-law model\footnote{\texttt{TBABS$_{\mathrm{Gal}}\times$(ZTBABS$\times$ZPOW)}}. The values of $\Gamma$ and $N_{\rm\,H}$ obtained for the three different spectra are consistent within their uncertainties, and we can therefore exclude that different flux levels are related to changes of the neutral absorber. In Fig.\,\ref{fig:ratiofluxstates} we illustrate the ratio between the data and the absorbed power-law model for each of the three spectra. The figure shows that the strength of the excess around 6\,keV decreases for increasing values of the flux.

We fitted the three spectra with the best absorption model discussed in Sect.\,\ref{sect:specan2} and shown in Panel\,D of Fig.\,\ref{fig:ratio_datamodels} (power law obscured by two ionized absorbers plus reflection from neutral material, a blackbody component, and four narrow emission lines representing Fe K$\alpha$, Fe K$\beta$, Fe\,\textsc{xxv} and Fe\,\textsc{xxvi}), fixing all the parameters to the values obtained by fitting the average {\it Suzaku} spectrum, leaving only the normalization of the power-law continuum free to vary. The fits yield good values of the reduced chi-squared: $\chi^2$/DOF=123.2/127, 428.5/431 and 248.2/255 for the low, mean and high flux state, respectively. The fit to the spectra is represented by the red dashed lines in Fig.\,\ref{fig:ratiofluxstates} (top panel). This is consistent with a scenario in which only the X-ray continuum is varying, while the ionized absorbers and the four emission lines are constant, as expected in the case of a distant reflector that does not respond simultaneously to changes of the X-ray continuum emission. This is different from the model proposed by \cite{Miyakawa:2012vn}, which explains X-ray variability as being solely due to variations of the covering factor of the ionized absorbers.

We then fitted the three spectra with the best broad Fe K$\alpha$ line model presented in Sect.\,\ref{sect:broadline} and shown in Panel\,F of Fig.\,\ref{fig:ratio_datamodels} (absorbed power-law plus broad Fe K$\alpha$ line and narrow Fe K$\alpha$ and Fe K$\beta$ lines), fixing all the parameters and allowing only the normalization of the power-law continuum to vary. 
Also in this case we obtained good values of the reduced chi-squared: $\chi^2$/DOF=122.1/127, 428.2/431 and 248.2/255 for the low, mean and high flux state, respectively (bottom panel of Fig.\,\ref{fig:ratiofluxstates}). The EW of the broad line is measured as in Sect.\,\ref{sect:specan_abs}, and it decreases for increasing values of the flux, and is $1070\pm30$\,eV, $787\pm12$\,eV, $620\pm11$\,eV for the low, mean and high flux state, respectively. The uncertainties associated to these values are lower than those reported in Table\,\ref{tab:resultsFit_broadline} because most of the parameters were frozen to their average values. The flux of the broad line in the three observations is constant ($f_{\rm\,K\alpha}\simeq 1.9\times10^{-13}\rm\,erg\,cm^{-2}\,s^{-1}$), and corresponds to a luminosity of $L_{\rm\,K\alpha}\simeq2.2\times10^{42}\rm\,erg\,s^{-1}$.
The time-scale we are probing ($\sim 6$\,ks) is longer than the time delay usually found between the broad Fe K$\alpha$ and the X-ray continuum ($\sim0.05-2$\,ks for black hole masses $M_{\rm\,BH}\lesssim 10^8 M_\odot$, e.g., \citealp{Zoghbi:2012uq,Zoghbi:2013fk,Kara:2013fk}). Therefore, unless the black hole is extremely massive, we can conclude that the constant flux of the broad Fe K$\alpha$ line is consistent with a scenario in which the reflecting material is not responding to the variations of the primary X-ray emission. This is in agreement with the idea that light-bending is at work in IRAS\,00521$-$7054, as hinted also by the large value of the reflection parameter.
Leaving the parameters of the broad line ($q$, $a$, $i$ and normalisation) and of the primary X-ray emission ($\Gamma$ and $N_{\rm\,H}$) free improves only marginally the fits ($\Delta \chi^2 =$2.6, 2.5 and 1.5 for the low, mean and high flux state, respectively), and we obtained that the three spectra have parameters consistent within their uncertainties. 

\section{Discussion and conclusions}

In this paper we studied the X-ray spectrum of IRAS\,00521$-$7054, the first type-II AGN in which the presence of an extremely large Fe\,K$\alpha$ line has been claimed, using new {\it Suzaku} data and archival {\it XMM-Newton} observations. We confirmed the existence of a strong excess at $\sim 6$\,keV above the power-law continuum, the largest ever observed in a Seyfert\,2, and found that this feature could be explained by two different scenarios.

In the {\it absorption scenario} (Sect.\,\ref{sect:specan2}) the X-ray power-law continuum is obscured by two layers of ionized material with similar column densities but different values of the ionization parameter. Besides this primary X-ray emission, the spectra require reflection from neutral material, possibly associated with the molecular torus, a blackbody component and four emission lines corresponding to Fe K$\alpha$, Fe K$\beta$, Fe\,\textsc{xxv} and Fe\,\textsc{xxvi}.

In the {\it reflection scenario} (Sect.\,\ref{sect:specan3}) the X-ray source is obscured by neutral material ($\log N_{\rm\,H}\sim 22.9$), and shows a strong ($R=2.7^{+2.5}_{-1.3}$) blurred reflection component arising from an ionized disk around a rotating SMBH with a spin $a \geq 0.73$. Two narrow lines representing emission from Fe K$\alpha$ and Fe K$\beta$ are also needed. The EW of the narrow Fe K$\alpha$ line is consistent with the value expected for an AGN with a luminosity of $2.5\times 10^{43}\rm\,erg\,s^{-1}$ in the 2--10\,keV band (\citealp{Bianchi:2007vn}, \citealp{Ricci:2013ij,Ricci:2013uq}, see Fig.\,7 of \citealp{Ricci:2013ij}). 
Our analysis shows that the reflecting material has a super-solar iron abundance ($Z_{\rm\,Fe}=1.6-4.7\,Z_\odot$). This is consistent with what has been found by several other studies of this type carried out on objects showing broad Fe K$\alpha$ lines (e.g., \citealp{Fabian:2012ve}), although these previous studies used \texttt{REFLIONX}, which could over-predict the Fe abundance by up to a factor of two \citep{Garcia:2014tg}. An explanation to the high Fe abundance has been recently put forward by \cite{Reynolds:2012ly}, who proposed that the radiative levitation of iron ions in the innermost part of the accretion flow disks could increase the photospheric abundance of iron.
In a recent work, \cite{Patrick:2012qf} systematically studied broad Fe K$\alpha$ lines by using {\it Suzaku} observations for a sample of 46\,AGN. From their study they found an average emissivity index of the accretion disk of $q=2.4\pm0.1$, lower than the value obtained for IRAS\,00521$-$7054 ($q=3.1^{+0.5}_{-0.4}$). The value of $q$ we obtained is consistent with a scenario in which a flat disk is illuminated by a central point source.
\cite{Patrick:2012qf} found an average inclination angle with respect to the accretion disk of $i=33^{\circ}\pm2^{\circ}$ for their sample of type-I AGN, while \cite{de-La-Calle-Perez:2010fk}, studying {\it XMM-Newton} observations of bright Seyfert\,1s, obtained $i=28^{\circ}\pm5^{\circ}$. Therefore it would seem that our results ($i = 40^{+2}_{-3}\rm\,deg$) are consistent with IRAS\,00521$-$7054 having a larger inclination angle than Seyfert\,1s, as predicted by the unification model (e.g., \citealp{Antonucci:1993kb}). Interestingly, the inclination angle of IRAS\,00521$-$7054 is consistent with that found for the obscured NLS1 NGC\,5506 ($i \simeq 40^{\circ}$, \citealp{Guainazzi:2010kx}).
\cite{Patrick:2012qf} found that the average equivalent width of the broad lines of their sample is $97\pm19$\,eV (Fig.\,9 of their paper), with the strongest one being that associated to Mrk\,79 ($EW=199^{+40}_{-37}\rm\,eV$). With an equivalent width of $EW=798^{+152}_{-102}$\,eV, the broad feature of the Seyfert\,2 IRAS\,00521$-$7054 is one of the strongest detected so far in AGN, comparable to those observed in the narrow line Seyfert\,1s IRAS\,13224$-$3809 \citep{Boller:2003qf,Fabian:2013vn} and 1H\,0707$-$495 \citep{Fabian:2009zr}, and to the low-flux state of MCG$-6-30-15$ (e.g., \citealp{Fabian:2003dq}). As discussed in Sect.\,\ref{sect:lc}, on the time-scale of a few years and of one month the source does not seem to vary, as the flux level of the {\it Suzaku} observation is the same as that measured by the two {\it XMM-Newton} observations carried out in March and April 2006. This seems to exclude that the source was in a low-flux state at the time of the {\it Suzaku} observation, which could account for the very large EW observed. Therefore, IRAS\,00521$-$7054 might be one of the very few obscured AGN to show evidence of strong light-bending.

In both scenarios we found that the energy centroid of the Fe K$\beta$ line, and the ratio between the Fe K$\beta$ and K$\alpha$ intensities, are consistent with the narrow iron lines being produced by ionized iron. The ionization state of the iron is Fe\,\textsc{xiv}--Fe\,\textsc{xvi} for the ionized absorption model, while in the blurred reflection scenario is Fe\,\textsc{xii}--Fe\,\textsc{xvi}, although in this case the presence of a Fe K$\beta$ line is only marginally significant. The material producing these narrow features might therefore be associated with the outer skin of the molecular torus or with the BLR.

While the X-ray continuum is variable in the time-scale of a few kiloseconds, the broad feature at $\sim 6$\,keV appears to be constant on a time scale of $\sim 6\rm\,ks$.
This lack of variability is consistent with both scenarios (Sect.\,\ref{sect:timeresolved}); for the ionized absorption model it fits the idea that the reprocessing material is associated with distant material, while in the case of the reflection scenario this would point towards light-bending, consistently with the fact that the reflection parameter is larger than one.

In both scenarios we obtained a steep power-law emission, with a photon index of $\Gamma\simeq 2.2-2.3$ (Column\,3 of tables \ref{tab:resultsFit_abs} and \ref{tab:resultsFit_broadline}), consistent with the value reported by \cite{Tan:2012fk}, and significantly larger than the average value of type-II AGN ($\Gamma\sim 1.8-1.9$, \citealp{Beckmann:2009fk,Ricci:2011fk}). Given the positive correlation between $\Gamma$ and the Eddington ratio ($\lambda_{\rm\,Edd}$, e.g. \citealp{Shemmer:2008fk}), this might imply that the source is accreting at higher values of $\lambda_{\rm\,Edd}$ than typical type-II AGN.
Spectropolarimetric studies did not find any evidence of a broad H$\alpha$ line \citep{Young:1996cr}, although \cite{Tran:2001oq,Tran:2003nx} erroneously reported IRAS\,00521$-$7054 as a source showing broad lines in polarised light. The presence of polarised broad lines would exclude that the source is an obscured NLS1. NLS1s tend to have lower black hole masses and accrete at higher values of $\lambda_{\rm\,Edd}$ than type-I AGN, which leads them to show steeper primary X-ray emission. NLS1s also usually show strong reflection features from the innermost part of the accretion flow (e.g., \citealp{Gallo:2006hc,Fabian:2009zr}). NGC\,5506 was also classified as a type-II AGN with polarised broad lines \citep{Tran:2001oq,Tran:2003nx}, while this object was later found, by using near-IR spectroscopy, to be the first obscured NLS1 \citep{Nagar:2002ly}. 
NLS1s are typically identified based on their optical properties: they show Fe\,II emission lines, a ratio $[\mathrm{OIII}]/\mathrm{H}\beta<3$ and FWHM$(\mathrm{H}\beta)<2000\rm\,km\,s^{-1}$ (e.g., \citealp{Komossa:2006uq} and references therein). Due to the obscuring material in the line of sight we cannot infer the properties of the BLR, so that it is not possible to verify these criteria for IRAS\,00521$-$7054. Considering that we can observe only the component of the $\mathrm{H}\beta$ produced in the narrow line region (NLR), which gives us a lower limit on the total $\mathrm{H}\beta$ flux, we can use the observed ratio $[\mathrm{OIII}]/\mathrm{H}\beta(\mathrm{NLR})=10$ \citep{Yu:2011kx} as an upper limit on the value of $[\mathrm{OIII}]/\mathrm{H}\beta$. This does not allow us to confirm the NLS1 nature of IRAS\,00521$-$7054, and more accurate near-IR and spectropolarimetric measurements are required to verify this possibility. So far only NGC\,5506 has been confirmed to be an obscured NLS1, although, two other Seyfert\,2s (NGC\,7314 and NGC\,7582) have been proposed to belong to this class \citep{Dewangan:2005tg}. IRAS\,00521$-$7054 could therefore be the obscured counterparts of objects such as IRAS\,13224$-$3809 and 1H\,0707$-$495, although it lacks the sharp decrease in flux at 7\,keV observed in these sources. Alternatively the source might be an intermediate object, i.e. a Seyfert\,2 with a high accretion rate. However, if the source is a genuine true Seyfert\,2 (i.e., an AGN without a BLR), this hypothesis would be in conflict with the prediction that the lack of a BLR is due to low values of the accretion rate (e.g., \citealp{Nicastro:2000uq}). If the Seyfert\,2 nature of IRAS\,00521$-$7054 were to be confirmed, this object would be the first type-II AGN to show evidence of strong light-bending.

The two scenarios discussed above to explain the X-ray spectrum are statistically indistinguishable, and future observations of IRAS\,00521$-$7054 will help shedding light on this very interesting source. In particular, with the advent of the high-resolution X-ray calorimeter on board {\it ASTRO-H} \citep{Takahashi:2010uq}, it will possible to clearly detect absorption lines from ionized outflows in the Fe K$\alpha$ region, and to break the degeneracy between absorption and reflection models.

\acknowledgments

We thank the anonymous referee for his/her comments, that helped us to improve the quality of our manuscript. We are grateful to M. Guainazzi, L. Gallo and F. Tombesi for useful discussion, to K. Ebisawa for kindly providing us with his XSTAR table model, to Y. Fukazawa for the new tuned {\it Suzaku} HXD/PIN background files and to Chin-Shin Chang and T. Kawamuro for their comments on the manuscript.
This research has made use of data obtained from the {\it Suzaku} satellite, a collaborative mission between the space agencies of Japan (JAXA) and USA (NASA), and from {\it XMM-Newton}, an ESA science mission with instruments and contributions directly funded by ESA Member States and NASA.
CR is a Fellow of the Japan Society for the Promotion of Science (JSPS). This work was partly supported by the Grant-in-Aid for Scientific Research 26400228 (YU) from the Ministry of Education, Culture, Sports, Science and Technology of Japan (MEXT). This research has made use of the NASA/IPAC Extragalactic Database (NED) which is operated by the Jet Propulsion Laboratory, of data obtained from the High Energy Astrophysics Science Archive Research Center (HEASARC), provided by NASA's Goddard Space Flight Center, and of the SIMBAD Astronomical Database which is operated by the Centre de Donn\'ees astronomiques de Strasbourg.

{\it Facilities:} \facility{Suzaku}, \facility{XMM-Newton}.


\bibliographystyle{apj} 
\bibliography{IRAS.bib}

\end{document}